\documentclass{PoS}
\usepackage{xspace}

\newcommand*{\ttbar}{\ensuremath{t \bar{t}}\xspace}
\newcommand*{\TeV}{\ifmmode {\mathrm{\ Te\kern -0.1em V}}\else
                   \textrm{Te\kern -0.1em V}\fi}%
\newcommand*{\GeV}{\ifmmode {\mathrm{\ Ge\kern -0.1em V}}\else
                   \textrm{Ge\kern -0.1em V}\fi}%
\newcommand*{\MeV}{\ifmmode {\mathrm{\ Me\kern -0.1em V}}\else
                   \textrm{Me\kern -0.1em V}\fi}%
\let\tev=\TeV
\let\gev=\GeV
\let\mev=\MeV
\def\ifb{\mbox{fb$^{-1}$}}

\title{Top-quark mass measurements at the LHC: alternative methods}

\ShortTitle{Top-quark mass measurements at the LHC: alternative methods}

\author{\speaker{Marcel Vos}%
        , on behalf of the ATLAS and CMS collaborations \\
        IFIC (UVEG/CSIC) Valencia\\
        Apartado de Correos 22085\\
        E-46071 Valencia - Spain\\ 
        E-mail: \email{Marcel.Vos@ific.uv.es}}

\abstract{Alternative top quark mass determinations can provide inputs to the world average with orthogonal systematic uncertainties and may help to refine the interpretation of the standard method. Among a number of recent results I focus on the extractions by ATLAS and CMS of the top quark pole mass from the \ttbar{} pair and \ttbar{} + 1 jet production cross-section, which have now reached a precision of 1\%.}

\FullConference{8th International Workshop on Top Quark Physics\\
		 14-18 September, 2015\\
		 Ischia, Italy}

\begin{document}

\section{Introduction}

The top quark mass is one of the key parameters of the Standard Model (SM).
The SM predicts a relation between the top quark mass and other parameters
of the theory (in particular the masses of the W and Higgs bosons). The top 
quark mass is the main input to extrapolations of the Higgs potential
to high scales. A precise measurement of the top quark mass
thus allows for a stringent test of the self-consistency of the theory. 
A precise measurement is furthermore needed to reduce the 
parametric uncertainties on many SM predictions.
 
Contrary to other quarks, the top quark decays before it hadronizes.
One can therefore observe a mass peak directly by reconstructing the products of the decay chain $ t \rightarrow Wb \rightarrow q\bar{q}b \rightarrow
$ jets of hadrons. As the partonic final state contains coloured objects,
observables describing the final state objects receive 
corrections from the parton shower and non-perturbative effects. 
The most precise measurements of the top quark mass extract
the mass by comparing distributions generated using Monte Carlo (MC) 
generators to the data. The mass parameter is usually identified with 
the pole mass. Such {\em direct} measurements of the top quark mass 
have attained a precision of better than 0.5\%~\cite{ATLAS:2014wva}, 
already exceeding the pre-LHC expectations~\cite{Beneke:2000hk}. 
 The current world average is dominated by 
systematic uncertainties on the response to jets of the experiments and 
in the modelling of the \ttbar{} signal.
Further progress is envisaged such that the precision could reach 
500~\mev{}~\cite{Juste:2013dsa} or even 
200~\mev{}~\cite{CMS-PAS-FTR-13-017} after the complete LHC programme.
This standard approach is the subject of several other 
contributions in these proceedings~\cite{castro,deterre,maier}.

The challenges inherent in performing a per mil level quark mass 
measurement and in the interpretation of the measured mass within a 
field-theoretical mass scheme~\cite{Juste:2013dsa,Moch:2014tta}
are addressed from several complementary angles:
\begin{itemize}
\item Estimates of non-perturbative effects and their uncertainties
are continuously being refined and must evolve beyond the traditional comparison of several parton shower models. 
\item Dedicated studies are ongoing to understand the (hopefully universal) 
relation of the MC mass parameter with the pole mass~\cite{ahoang14,corcella}.
\item Measurements that base the top quark measurement on 
alternative observables, with an orthogonal dependence 
on the main sources systematics, strengthen the world average.
\item The top quark pole mass is extracted from a corrected measurement of the
(differential) cross-section. 
\end{itemize}

For a discussion of the first and second approaches, the reader is
referred to other contributions to these 
proceedings~\cite{corcella}. 

Good examples of the third strategy are the methods which use the 
relation of the top quark mass with the B-hadron decay 
length~\cite{Hill:2005zy} or the invariant mass of the
$J/\psi$-lepton system~\cite{Kharchilava:1999yj,Chierici:951386}. 
These methods, that are promising 
with large integrated luminosity, have been explored by 
the ATLAS and CMS experiments: Reference~\cite{CMS-PAS-TOP-12-030}
determines the mass to a precision of 3~\gev{} using the
B-hadron decay length, while 
References~\cite{ATLAS-CONF-2015-040} and~\cite{CMS-PAS-TOP-13-007} investigate
the selection and reconstruction of $J/\psi$ in \ttbar{} events.
Two further methods that rely on relations of the top quark mass with 
certain features (peaks, end-points) of observables of the top quark decay
products that have recently been deployed by the CMS experiment are 
presented in these proceedings: the top quark mass measurement from
the $b$-jet energy peak and from the invariant mass $m_{bl}$ of the system 
formed by the $b$-jet and charged lepton from top quark decay.

The extraction of the top quark mass from the production cross-section 
forms an independent cross-check of the standard interpretation of the
 {\em direct} measurement in the pole mass scheme. 
Recent extractions of the top quark pole mass
from the total cross-section are presented in Section~\ref{sec:xsec}, 
while the analysis of the differential \ttbar{} + 
1 jet cross-section is presented in Section~\ref{sec:diffxsec}.

\section{Top quark mass measurement from the b-jet energy peak}
\label{sec:epeak}

In the rest frame of the top quark the energy of the b-quark produced in
the $t\rightarrow Wb$ decay has a simple relation with the top quark mass.
In Reference~\cite{Agashe:2012bn} the authors show that under certain 
assumptions (in particular that of unpolarized production of the top quarks) 
the same relation holds for the peak position of the b-jet energy
in the laboratory frame. Subsequent papers study how this observation
can be used to measure the mass of new particles~\cite{Agashe:2013eba} and 
work generalize the study to multi-body decays~\cite{Agashe:2015wwa}.

CMS has applied this technique to the 8~\tev{} proton-proton collision data 
collected in 2012~\cite{CMS-PAS-TOP-15-002}. A very pure $t\bar{t}$ sample, 
with an expected \ttbar{} contribution of 88\% for the single b-tag
and 95\% for the double b-tag category, is selected in the 
di-lepton channel with an isolated electron and muon of opposite electric 
charge. The peak position in the $b-$ jet energy distribution is
found with a fit to a $1/E \log{E}$ function and calibrated to the
parton-level b-quark energy, after which the top quark mass is found from 
the simple relation $m_t = E_b + \sqrt{m^2_W - m^2_b + E^2_b}$. 
The precision of the measurement is limited by uncertainties in the jet 
energy scale (1.2~\gev{}) and in \ttbar{} modelling (2.1~\gev{}). The result,
$m_t = $ 172.3 $\pm$ 2.9~\gev{}, is in excellent agreement with
the world average.

\section{Template fit to the $m_{bl}$ distribution}
\label{sec:mbl}

The second alternative mass determination I discuss measures the top
quark mass from the distribution of the invariant mass of the system 
formed by the b-jet and the charged lepton. 
Following Ref.~\cite{Biswas:2010sa} the lepton and b-jet are combined
that yield the minimal mass. This distribution is known to be  
invariant under Lorentz boosts. The extracted top quark mass is
therefore expected to be insensitive to the production mechanism.
The shape of the distribution depends on the top quark mass,
with the most pronounced sensitivity around the end-point of
the distribution (for $m_{lb}^{min} = \sqrt{m_{t}^2 - m_{W}^2}$).

In Ref.~\cite{CMS-PAS-TOP-14-014} the CMS collaboration reconstructs 
the $m_{lb}^{min}$ distribution in the very pure electron-muon channel,
using the full 2012 data set (20~\ifb at $\sqrt{s}=$ 8~\tev). The top
quark mass is determined by fitting the distribution to templates
created with the MadGraph Monte Carlo generator. The mass
is determined from the absolute and normalized distribution, with the
shape analysis yielding more powerful results. The result is
$m_t = $ 172.3 $\pm$ 1.3~\gev{}, with the dominant uncertainties
in the modelling of top quark production and decay.

The mass extraction is repeated with templates based on a fixed-order 
prediction implemented in MCFM that describes top quark pair production 
at NLO and has leading-order accuracy for the decay. The parton-level 
distribution is ``forward folded'' to the detector-level (i.e. a MC estimate
of the effects of hadronization and detector resolution is applied
in the form of a smearing matrix and bin-by-bin efficiency correction).
This is a promising step towards an extraction of the pole 
mass from the $m_{lb}^{min}$ distribution. The relatively large shift
of the result (900~\mev{}, much larger than the scale uncertainty of 100~\mev{}
assigned to the MCFM prediction) demonstrates that for a pole mass extraction 
with reliable uncertainties the top quark decay must 
be incorporated at NLO in the calculation. Several 
authors~\cite{Heinrich:2013qaa} 
have pointed out that especially in the end-point region 
NLO decay and off-shell effects are sizeable.

\section{Extraction from the total \ttbar{} production cross-section}
\label{sec:xsec}

The classical alternative top mass measurement is the extraction 
of the mass from the total top quark pair production cross-section, 
first performed in 
Ref.~\cite{Langenfeld:2009wd,Abazov:2009ae}. The mass is
inferred by comparing the measured cross-section to a precise
prediction of the dependence of the inclusive
\ttbar{} cross-section on the top quark mass. 
Among the advantages of this method is the possibility to unambiguously 
choose the renormalization scheme used in the calculation. This feature
has been used to extract the running $\bar{MS}$ mass 
directly~\cite{Langenfeld:2009wd,Abazov:2011pta}.
The theoretical uncertainty due to the truncation of the perturbative series 
is readily evaluated in the usual way, by varying the renormalization and
factorization scales.

The top quark mass is extracted by maximizing the product of the likelihoods 
corresponding to the measured cross-section at center-of-mass energies 
of 7 and 8~\tev{}. Experimental and theoretical uncertainties are treated 
as nuisance parameters, taking into account the strong correlations between
the 7 and 8~\tev{} measurements. 
The most recent ATLAS and CMS measurements use
the NNLO calculation of Ref.~\cite{Czakon:2013goa} with NNLL resummation
to extract the pole mass, which reduces the scale uncertainty on the
cross-section to the level of 3\%. An uncertainty of 1.7-1.8\% is assigned
to the cross-section to account for the uncertainty in the LHC beam energy.

The ATLAS analysis of Ref.~\cite{Aad:2014kva} extracts the pole mass for
7~\tev{} and 8~\tev{} separately, and for three exponents of 
the previous generation of PDF sets (CT10, MSTW2008 and NNPDF2.1).
The results obtained with different PDF sets are compatible with each 
other to within 200~\mev{}, but the results obtained on 7~\tev{} and 8~\tev{} 
data cluster around 171.4~\gev{} and 174.1~\gev, respectively.
Considering only uncorrelated experimental uncertainties, the two values 
are consistent at the level of 1.7 standard deviations.
The combined fit yields 172.9 $^{+2.5}_{-2.6}$~\gev{}. The uncertainty
is dominated by the PDF uncertainty, estimated as the full envelope 
of the error sets of the three PDFs, which amounts to 1.8~\gev{}. 

The preliminary result of the cross-section measurement by CMS in the 
$e \mu$ channel analysis of the 7 and 8~\tev{} 
data~\cite{CMS-PAS-TOP-13-004}, presented in detail elsewhere in these 
proceedings~\cite{crucy}, is slightly more precise than the ATLAS 
result~\cite{glatzer,Aad:2014kva}: where ATLAS finds 
an uncertainty of 3.9\% (4.3\%) at $\sqrt{s}=$ 7~\tev{} ($\sqrt{s}=$ 8~\tev{}) 
the CMS uncertainty is 3.6\% (3.9\%). Importantly, in both measurements
the dependence of the measured cross-section on the assumed top quark mass 
is negligible~\footnote{In previous analyses~\cite{Langenfeld:2009wd,Abazov:2011pta,Chatrchyan:2013haa} the straightforward interpretation of the extracted
mass was somewhat obfuscated by this dependence.}.
 
CMS uses the most recent PDF sets in the mass 
extraction, that include constraints from LHC data\footnote{The MMHT14 and 
NNPDF3.0 sets include the LHC \ttbar{} production cross-section measurements 
as a constraint. Even if the bias on the mass due to the circularity of the 
exercise is expected to be negligible at present, an effort must be made to 
improve the PDFs without sacrificing a key observable in this mass 
determination and many constraints on physics beyond the Standard Model. } 
and quotes several results that assume a given PDF set and $\alpha_s$ value. 
The mass values extracted from 7 and 8~\tev{} data agree within 500~\mev{}.
The values obtained with three different PDF sets span 500-600~\mev{} and
have not been combined into a single mass value~\footnote{The PDF4LHC 
recommendation~\cite{Butterworth:2015oua} for run II explicitly requests
top quark mass extraction using individual PDF sets, as provided by 
both collaborations. The new recommendation for obtaining the total PDF
uncertainty using the combined PDF set is less conservative than the
envelope.}.

In Table~\ref{tab:polemass} these results are compared to the ATLAS results
and the other measurements of the top quark pole mass. The results are
in good agreement with each other and with the world average from
the {\em direct} measurements.

\begin{table}
\caption{The top quark pole mass extracted from the production cross-section.}
\begin{tabular}{l|c|c|c|c}
Experiment    & pole mass & data & theory & comment \\ \hline 
D0~\cite{Abazov:2009ae}        & 169.1$^{+5.9}_{-5.2}$~\gev{}  & $p\bar{p}$, 1.96~\tev{}, 1~\ifb{}   &  \\
Langenfeld             & 168.9$^{+3.5}_{-3.4}$~\gev{}  & idem   & through  \\
et al.~\cite{Langenfeld:2009wd}  &  &  & $\bar{MS}$ mass \\ 
D0~\cite{Abazov:2011pta}       & 167.5 $^{+5.4}_{-4.9}$~\gev{} & $p\bar{p}$, 1.96~\tev{}, 5.3~\ifb{}          & approx NNLO & \\
CMS~\cite{Chatrchyan:2013haa}  & 176.7 $\pm$ 2.9~\gev{}     & $pp$, 7~\tev{}, 5~\ifb{}  & approx NNLO & \\
ATLAS~\cite{Aad:2014kva}       & 172.9 $^{+2.5}_{-2.6}$~\gev{} & $pp$, 7~\tev{}, 5~\ifb{}   &  NNLO+NNLL & full PDF4LHC~\cite{Rojo:2015acz} \\
                               &                             &  + 8~\tev{}, 20~\ifb{}  & &  envelope of 3 PDF sets \\
CMS~\cite{CMS-PAS-TOP-13-004}  & 173.6 $^{+1.7}_{-1.8}$~\gev{} &  idem & idem & NNPDF3.0~\cite{Ball:2014uwa}, preliminary\\
CMS~\cite{CMS-PAS-TOP-13-004}  & 173.9 $^{+1.8}_{-1.9}$~\gev{} &  idem & idem & MMHT2014, preliminary \\
CMS~\cite{CMS-PAS-TOP-13-004}  & 174.1 $^{+2.1}_{-2.2}$~\gev{} &  idem & idem & CT14~\cite{Dulat:2015mca}, preliminary \\ \hline
ATLAS~\cite{Aad:2015waa}       & 173.7 $^{+2.3}_{-2.1}$~\gev{} & $pp$, 7~\tev{}, 5~\ifb{}   &  NLO \ttbar{} + 1 jet & full PDF4LHC~\cite{Rojo:2015acz} \\
\end{tabular}
\label{tab:polemass}
\end{table}

\section{Extraction from the differential \ttbar{} + 1 jet cross-section}
\label{sec:diffxsec}

A method proposed in Ref.~\cite{Alioli:2013mxa} extracts the pole mass from
the differential cross-section in top quark pair production in association
with a hard jet ($d \sigma / d m_{t\bar{t} j}$, with $m_{t\bar{t}j}$ 
the invariant mass of the system formed by the top quark pair and the 
extra jet). The enhanced sensitivity to the top quark mass avoids the 
limit on the precision of the quark pole mass extraction from the 
theory uncertainty on the inclusive \ttbar{} production cross-section. 

An ATLAS measurement~\cite{Aad:2015waa} of the pole mass using this method 
on the $\sqrt{s}=$ 7~\tev{} data set collected in 2011 yields a result 
of $m_t^{pole} = $ 173.7 $\pm$ 1.5 (stat.) $\pm$ 1.4 (syst.) 
$^{+1.0}_{-0.5}$~\gev{}, where the latter term represents the 
theory uncertainty, estimated from variations of the renormalization
and factorization scales and the envelope of the error sets from PDF fits.

\section{Summary and prospects}
\label{sec:summary}

The key goal of alternative top quark mass measurement is to provide an
independent confirmation of the interpretation of the more precise 
standard method. With the publication of the results of the pole mass 
extraction from the inclusive top quark pair production cross-section
measured at 7 and 8~\tev{}~\cite{Aad:2014kva,CMS-PAS-TOP-13-004} and 
from the differential $\ttbar{} + 1$ jet cross-section at 
7~\tev~{}~\cite{Aad:2015waa} the precision of such measurements 
approaches 1\%. To that level of precision the results are in
excellent agreement with the world average.

Further progress can be made by combining the 
ATLAS and CMS measurements, by reducing systematics on the 13~\tev{} data
and by improving the PDF fit by including LHC data.
A significant reduction of the weight of the PDF uncertainty is expected
at $\sqrt{s}=$ 13~\tev{}, where top quark pair production requires a 
smaller fraction $x$ of the proton momentum. If both the experimental and
PDF uncertainties are reduced considerably, the precision of the mass
extraction is limited by the scale uncertainty of the NNLO+NNLL calculation~\footnote{The scale variation of the NNLO calculation in the $\bar{MS}$ scheme
is about half that of the NNLO+NNLL calculation in the pole mass scheme,
but the consensus is that in this case the uncertainty may be underestimated.}
to approximately 1~\gev{}. 

The extraction from the differential 
cross-section~\cite{Aad:2015waa} can improve considerably by including
the 8~\tev{} (and 13~\tev) data. With a finer-grained binning the 
sensitivity to the top quark mass increases considerably. The authors
of Ref.~\cite{Alioli:2013mxa} expect a 1~\gev{} precision can be achieved
with the data set collected in 2012 and 2015.

Methods like the $m_{bl}$ analysis performed by CMS~\cite{CMS-PAS-TOP-14-014} 
have shown good sensitivity to the top quark mass. 
With a more sophisticated (NLO) 
treatment of the modeling of the top quark decay and a careful evaluation
of the MC mass dependence of the unfolding and of the theory uncertainty 
such measurements can form valuable pole mass measurements.

\bibliographystyle{JHEP}
\bibliography{confnotebib_atlas}

\providecommand{\href}[2]{#2}\begingroup\raggedright\begin{thebibliography}{10}

\bibitem{ATLAS:2014wva}
{\bf ATLAS, CDF, CMS, D0} Collaboration, {\it {First combination of Tevatron
  and LHC measurements of the top-quark mass}},
  \href{http://arxiv.org/abs/1403.4427}{{\tt arXiv:1403.4427}}.

\bibitem{Beneke:2000hk}
M.~Beneke et~al., {\it {Top quark physics}},  in {\em {1999 CERN Workshop on
  standard model physics (and more) at the LHC, CERN, Geneva, Switzerland,
  25-26 May: Proceedings}}, 2000.
\newblock \href{http://arxiv.org/abs/hep-ph/0003033}{{\tt hep-ph/0003033}}.

\bibitem{Juste:2013dsa}
A.~Juste, S.~Mantry, A.~Mitov, A.~Penin, P.~Skands, E.~Varnes, M.~Vos, and
  S.~Wimpenny, {\it {Determination of the top quark mass circa 2013: methods,
  subtleties, perspectives}},  {\em Eur. Phys. J.} {\bf C74} (2014), no.~10
  3119, [\href{http://arxiv.org/abs/1310.0799}{{\tt arXiv:1310.0799}}].

\bibitem{CMS-PAS-FTR-13-017}
{\bf CMS} Collaboration, {\it {Projected improvement of the accuracy of
  top-quark mass measurements at the upgraded LHC}},  {\em CMS-PAS-FTR-13-017}.

\bibitem{castro}
A.~Castro, {\it {Top quark mass measurements at the LHC: standard methods}},
  in {\em {8th International Workshop on Top Quark Physics (TOP2015) Ischia,
  NA, Italy, September 14-18, 2015}}, 2015.

\bibitem{deterre}
C.~Deterre, {\it Top-quark mass measurements at the tevatron (incl. tevatron
  combination)},  {\em these proceedings} (2015).

\bibitem{maier}
A.~Maier, {\it Statistical and systematic treatment issues in top mass
  combinations},  {\em these proceedings} (2015).

\bibitem{Moch:2014tta}
S.~Moch et~al., {\it {High precision fundamental constants at the TeV scale}},
  \href{http://arxiv.org/abs/1405.4781}{{\tt arXiv:1405.4781}}.

\bibitem{ahoang14}
A.~H. Hoang, {\it {The Top Mass: Interpretation and Theoretical
  Uncertainties}},  \href{http://arxiv.org/abs/1412.3649}{{\tt
  arXiv:1412.3649}}.

\bibitem{corcella}
G.~Corcella, {\it {Interpretation of the Top-Quark Mass Measurements: A Theory
  Overview}},  in {\em {8th International Workshop on Top Quark Physics
  (TOP2015) Ischia, NA, Italy, September 14-18, 2015}}, 2015.
\newblock \href{http://arxiv.org/abs/1511.08429}{{\tt arXiv:1511.08429}}.

\bibitem{Hill:2005zy}
C.~S. Hill, J.~R. Incandela, and J.~M. Lamb, {\it {A Method for measurement of
  the top quark mass using the mean decay length of $b$ hadrons in $t \bar{t}$
  events}},  {\em Phys. Rev.} {\bf D71} (2005) 054029,
  [\href{http://arxiv.org/abs/hep-ex/0501043}{{\tt hep-ex/0501043}}].

\bibitem{Kharchilava:1999yj}
A.~Kharchilava, {\it {Top mass determination in leptonic final states with
  $J/\psi$}},  {\em Phys. Lett.} {\bf B476} (2000) 73--78,
  [\href{http://arxiv.org/abs/hep-ph/9912320}{{\tt hep-ph/9912320}}].

\bibitem{Chierici:951386}
R.~Chierici and A.~Dierlamm, {\it {Determination of the top mass with exclusive
  events t -> Wb -> l$\nu$ J/$\psi$ X}},  Tech. Rep. CMS-NOTE-2006-058, CERN,
  Geneva, May, 2006.

\bibitem{CMS-PAS-TOP-12-030}
{\bf CMS} Collaboration, {\it {Measurement of the top quark mass using the
  B-hadron lifetime technique}},  {\em CMS-PAS-TOP-12-030}.

\bibitem{ATLAS-CONF-2015-040}
{\bf ATLAS} Collaboration, {\it {Reconstruction of J/$\psi$ mesons in
  $t\bar{t}$ final states in proton-proton collisions at $\sqrt{s}$ = 8 TeV
  with the ATLAS detector}},  {\em ATLAS-CONF-2015-040}.

\bibitem{CMS-PAS-TOP-13-007}
{\bf CMS} Collaboration, {\it {Study of the underlying event, b-quark
  fragmentation and hadronization properties in t$\bar{t}$ events}},  {\em
  CMS-PAS-TOP-13-007}.

\bibitem{Agashe:2012bn}
K.~Agashe, R.~Franceschini, and D.~Kim, {\it {Simple 'invariance' of two-body
  decay kinematics}},  {\em Phys. Rev.} {\bf D88} (2013), no.~5 057701,
  [\href{http://arxiv.org/abs/1209.0772}{{\tt arXiv:1209.0772}}].

\bibitem{Agashe:2013eba}
K.~Agashe, R.~Franceschini, and D.~Kim, {\it {Using Energy Peaks to Measure New
  Particle Masses}},  {\em JHEP} {\bf 11} (2014) 059,
  [\href{http://arxiv.org/abs/1309.4776}{{\tt arXiv:1309.4776}}].

\bibitem{Agashe:2015wwa}
K.~Agashe, R.~Franceschini, D.~Kim, and K.~Wardlow, {\it {Mass Measurement
  Using Energy Spectra in Three-body Decays}},
  \href{http://arxiv.org/abs/1503.03836}{{\tt arXiv:1503.03836}}.

\bibitem{CMS-PAS-TOP-15-002}
{\bf CMS} Collaboration, {\it {Measurement of the top quark mass from the b jet
  energy spectrum}},  {\em CMS-PAS-TOP-15-002}.

\bibitem{Biswas:2010sa}
S.~Biswas, K.~Melnikov, and M.~Schulze, {\it {Next-to-leading order QCD effects
  and the top quark mass measurements at the LHC}},  {\em JHEP} {\bf 08} (2010)
  048, [\href{http://arxiv.org/abs/1006.0910}{{\tt arXiv:1006.0910}}].

\bibitem{CMS-PAS-TOP-14-014}
{\bf CMS} Collaboration, {\it {Determination of the top-quark mass from the
  m(lb) distribution in dileptonic ttbar events at sqrt(s) = 8 TeV}},  {\em
  CMS-PAS-TOP-14-014}.

\bibitem{Heinrich:2013qaa}
G.~Heinrich, A.~Maier, R.~Nisius, J.~Schlenk, and J.~Winter, {\it {NLO QCD
  corrections to $W^{+} W^{-}b\bar{b}$ production with leptonic decays in the
  light of top quark mass and asymmetry measurements}},  {\em JHEP} {\bf 06}
  (2014) 158, [\href{http://arxiv.org/abs/1312.6659}{{\tt arXiv:1312.6659}}].

\bibitem{Langenfeld:2009wd}
U.~Langenfeld, S.~Moch, and P.~Uwer, {\it {Measuring the running top-quark
  mass}},  {\em Phys.Rev.} {\bf D80} (2009) 054009,
  [\href{http://arxiv.org/abs/0906.5273}{{\tt arXiv:0906.5273}}].

\bibitem{Abazov:2009ae}
{\bf D0} Collaboration, V.~M. Abazov et~al., {\it {Combination of t anti-t
  cross section measurements and constraints on the mass of the top quark and
  its decays into charged Higgs bosons}},  {\em Phys. Rev.} {\bf D80} (2009)
  071102, [\href{http://arxiv.org/abs/0903.5525}{{\tt arXiv:0903.5525}}].

\bibitem{Abazov:2011pta}
{\bf D0} Collaboration, V.~M. Abazov et~al., {\it {Determination of the pole
  and MSbar masses of the top quark from the $t\bar{t}$ cross section}},  {\em
  Phys. Lett.} {\bf B703} (2011) 422--427,
  [\href{http://arxiv.org/abs/1104.2887}{{\tt arXiv:1104.2887}}].

\bibitem{Czakon:2013goa}
M.~Czakon, P.~Fiedler, and A.~Mitov, {\it {Total Top-Quark Pair-Production
  Cross Section at Hadron Colliders Through $O(\alpha_s^4)$}},  {\em Phys. Rev.
  Lett.} {\bf 110} (2013), no.~25 252004,
  [\href{http://arxiv.org/abs/1303.6254}{{\tt arXiv:1303.6254}}].

\bibitem{Aad:2014kva}
{\bf ATLAS} Collaboration, {\it {Measurement of the $t\overline{t}$ production
  cross-section using $e\mu $ events with $b$ -tagged jets in $pp$ collisions
  at $\sqrt{s}=7$ and 8 TeV with the ATLAS detector}},  {\em Eur. Phys. J.}
  {\bf C74} (2014), no.~10 3109, [\href{http://arxiv.org/abs/1406.5375}{{\tt
  arXiv:1406.5375}}].

\bibitem{CMS-PAS-TOP-13-004}
{\bf CMS} Collaboration, {\it {Measurement of the \ttbar{} production cross
  section in the $e\mu$ channel in pp collisions at 7 and 8 TeV}},  {\em
  CMS-PAS-TOP-13-004}.

\bibitem{crucy}
S.~Crucy, {\it Inclusive top pair production at 7, 8 and 13 tev in cms},  {\em
  these proceedings} (2015).

\bibitem{glatzer}
J.~Glatzer, {\it Inclusive top pair production at 7, 8 and 13 tev in atlas},
  {\em these proceedings} (2015).

\bibitem{Chatrchyan:2013haa}
{\bf CMS} Collaboration, {\it {Determination of the top-quark pole mass and
  strong coupling constant from the $t\bar{t}$ production cross section in pp
  collisions at $\sqrt{s}$ = 7 TeV}},  {\em Phys. Lett.} {\bf B728} (2014)
  496--517, [\href{http://arxiv.org/abs/1307.1907}{{\tt arXiv:1307.1907}}].
  [Erratum: Phys. Lett.B728,526(2014)].

\bibitem{Butterworth:2015oua}
J.~Butterworth et~al., {\it {PDF4LHC recommendations for LHC Run II}},
  \href{http://arxiv.org/abs/1510.03865}{{\tt arXiv:1510.03865}}.

\bibitem{Rojo:2015acz}
J.~Rojo et~al., {\it {The PDF4LHC report on PDFs and LHC data: Results from Run
  I and preparation for Run II}},  {\em J. Phys.} {\bf G42} (2015) 103103,
  [\href{http://arxiv.org/abs/1507.00556}{{\tt arXiv:1507.00556}}].

\bibitem{Ball:2014uwa}
{\bf NNPDF} Collaboration, R.~D. Ball et~al., {\it {Parton distributions for
  the LHC Run II}},  {\em JHEP} {\bf 04} (2015) 040,
  [\href{http://arxiv.org/abs/1410.8849}{{\tt arXiv:1410.8849}}].

\bibitem{Dulat:2015mca}
S.~Dulat et~al., {\it {The CT14 Global Analysis of Quantum Chromodynamics}},
  \href{http://arxiv.org/abs/1506.07443}{{\tt arXiv:1506.07443}}.

\bibitem{Aad:2015waa}
{\bf ATLAS} Collaboration, {\it {Determination of the top-quark pole mass using
  $ t\overline{t} $ + 1-jet events collected with the ATLAS experiment in 7 TeV
  pp collisions}},  {\em JHEP} {\bf 10} (2015) 121,
  [\href{http://arxiv.org/abs/1507.01769}{{\tt arXiv:1507.01769}}].

\bibitem{Alioli:2013mxa}
S.~Alioli et~al., {\it {A new observable to measure the top-quark mass at
  hadron colliders}},  {\em Eur. Phys. J.} {\bf C73} (2013) 2438,
  [\href{http://arxiv.org/abs/1303.6415}{{\tt arXiv:1303.6415}}].

\end{thebibliography}\endgroup

\end{document}